\documentclass[         %
aps,                    
prd,                    
showpacs,               
nofootinbib,            
showkeys,               %
preprintnumbers,        %
floatfix]               
{revtex4}               
\usepackage{graphicx,longtable}
\usepackage[dvips]{epsfig}
\usepackage{bm}
\usepackage{fancybox}

\begin{document}

\title{Single top quark production through unparticles at photon colliders }

\author{T. M. Aliev\footnote{Permanent address:
Institute of Physics, Baku, Azerbaijan}}
\email{taliev@metu.edu.tr}
\affiliation{Department of Physics, 
Faculty of Arts and Sciences, Middle East
Technical University, Ankara, Turkey}
\author{K. O. Ozansoy}
\affiliation{Department of Physics Engineering, 
Ankara University, 06100 Tandogan, Ankara, Turkey}
\email{oozansoy@science.ankara.edu.tr}
\date{\today}

\begin{abstract}
We investigate the single top quark production with
the exchange of unparticles through high
energy photon-photon collision $\gamma \gamma\to t \bar c$.
The effects of unparticles on the scattering cross section
for different polarization configurations, and
for various values of the scaling dimension $d, 1<d<2$,
is analysed. It is shown that the $(+-)$ polarisation
configuration is more preferable searching for unparticle physics
signatures.
\end{abstract}

\medskip


\maketitle


After launching the Large Hadron Collider(LHC) at CERN
new horizons in particle physics will be sought in
exploring new physics beyond the Standard Model(SM).
One of the research directions of the LHC is examining the
 predictions of the SM at the electroweak scale, as well as,
to discover possible new physics effects in details. For these goals,
the properties of the top quark which is the heaviest particle
in the framework of the SM will be studied comprehensively.
Being complementary to the LHC and
other high energy hadron colliders, multi-TeV $e^+e^-$ linear colliders,
such as the ILC(International Linear Collider)
and CLIC(Compact Linear Collider), which are free of difficulties
of the hadron contaminations effects, will be very powerful tools
to examine conceivible new physics outcomes. Comparing
with the hadron colliders, the $e^+e^-$ linear colliders
are very powerfull for precise determination of the masses
and spin properties of the potential new particles, \cite{clic}. Another
very important property of a linear $e^+e^-$ linear collider
is that it can be converted into  $e^-e^-, e^-\gamma$, or
$\gamma \gamma$ collider mode, where high energy photon beams are
generated by using the Compton backscattering of the initial electron
and laser photon beams. The energy and the luminosity of the photon
beams are practically same as of the initial electron and positron beams.
A detailed description of the photon collider is presented in \cite{Badelek:2001xb}.
Physics programme of the photon colliders is described in \cite{Telnov:2008zz}, and
in ~\cite{Ginzburg:1983}, a detailed analysis on
$\gamma\gamma$ option of an $e^+e^-$ collider has been given.
It should be noted that the cross sections for
the considered processes in $\gamma\gamma$ collisions are larger than
in the $e^+e^-$ case. Therefore, the photon colliders also can open new
windows  and new possibilities searching for new physics beyond the SM.

One of the new physics models beyond the SM is
the unparticle physics proposed by Georgi, \cite{Georgi:2007ek, Georgi:2007si}.
According to the Georgi`s scenario, if there is a conformal symmetry
in nature it should be broken at a very high energy scale
which must be above the current energy scale of the colliders.
Based on the idea of Banks and Zaks, \cite{Banks:1981nn}, Georgi
presents the scale invariant sector as a set of the
Banks-Zaks operators ${\cal O}_{BZ}$, and defines it at
the very high energy scale,  \cite{Georgi:2007ek}.
Interactions of the SM operators
${\cal O}_{SM}$ with the BZ operators ${\cal O}_{BZ}$
are expressed by the exchange of particles with a very high energy
mass scale ${\cal M}_{\cal U}^k$ as the following form

\begin{eqnarray}
\label{1}
 \frac{1}{{\cal M}_{\cal U}^k}{O}_{BZ}{O}_{SM}
\end{eqnarray}

where BZ, and SM operators are defined as
${O}_{BZ}\in {\cal O}_{BZ}$ with mass dimension $d_{BZ}$,
and ${O}_{SM} \in {\cal O}_{SM} $ with mass dimension $d_{SM}$.
Low energy effects of the scale invariant ${\cal O}_{BZ}$ fields
lead a  dimensional transmutation. Hence, after
the dimensional transmutation Eq.(\ref{1}) is given as

\begin{eqnarray}
\label{2}
 \frac{C_{\cal U} \Lambda_{\cal U}^{d_{BZ}-d}}
{{\cal M}_{\cal U}^k}{O}_{\cal U}{O}_{SM}
\end{eqnarray}

where $d$ is the scaling mass dimension(or anomalous dimension)
of the unparticle operator $O_{\cal U}$ ,
and the constant $C_{\cal U}$ is a coefficient function.
Interactions between the unparticles and the SM fields
have been listed by Ref~\cite{Cheung:2007ue}. Possible manifestations
of the unparticles via their direct and
indirect effects have been investigated in many works
( see for example, \cite{unparticle}, and references there in.).

In present work, we study the single top quark production in the
process $\gamma\gamma\to t \bar c$  in the $\gamma\gamma$ collider option of
a multi-TeV linear collider, eg. the CLIC, in unparticle physics.
Our results can easily be extended for other possible future
multi TeV-scale linear electron-positron colliders.

For the calculation of the matrix element of $\gamma \gamma \to t\bar c$ 
process in the unparticle model, the interaction vertices of unparticles 
with photon and quarks are needed. In this work, we consider only scalar
unparticle contribution, vector and tensor unparticle contributions
are neglected. The reason for neglecting the contributions of the 
vector and tensor unparticle contributions is as follows.
In \cite{Grinstein:2008qk}, it was shown that for the
vector and tensor unparticles $d>3$ and $d>4$ respectively.
Numerical calculations show that for these
values of d contributions of unparticles are negligible.


The effective interactions between the scalar unparticle and the
SM fermions, and the photons are given in the following form, respectively,

\begin{equation}
\frac{1}{\Lambda^{d-1}}\bar{f}(\lambda_{S}^{ff'}
+i\gamma_{5}\lambda_{P}^{ff'})f'O_{\cal U}
\label{eq:3a}
\end{equation}

\begin{equation}
\frac{1}{\Lambda^{d}} [\lambda_{0}F_{\mu\nu}F^{\mu\nu}
+\lambda_{0}^{\prime} \tilde{F}_{\mu\nu}F^{\mu\nu}]O_{\cal U}
\label{eq:3b}
\end{equation}

where $f$ and $f'$ denote different flavor of quarks, with
the same electric charge, $F_{\mu\nu}$ is the electromagnetic field
tensor, and
$\tilde{F}_{\mu\nu}=\frac{1}{2}\epsilon_{\mu\nu\alpha\beta}F^{\alpha\beta}$,
and $O_{\cal U}$ stands for the scalar 
unparticle( Ref.s~\cite{Cheung:2007ue}, \cite{Iltan:2008}).
Here we would like to make the following remark. In Ref.~
\cite{Fox:2007sy} it is obtained that interaction of the
unparticle sector with the SM fields can be proceeded via
interaction $~c_2H^{2}O_{\cal U}$, where $H$ is the Higgs field.
Hence this operator leads to the conformal symmetry breaking 
at low energy scales when the Higgs field gets the vacuum 
expectation value, and this symmetry breaking imposes some 
strong constranints on the unparticle sector.
Following the arguments present in  \cite{Fox:2007sy},
we assume $c_2<<1$. Under this condition
at TeV energy scale,  the effects of
unparticle sector on future high energy
collider energies can be probed. One can see from the above vertices
one of the very promising properties of the unparticle physics
is that it permits the excistence of the flavor changing neutral
current interactions in the interactions of the unparticles with the
SM particles at the tree level.

The scalar unparticle propagator is given as

\begin{eqnarray}
 \Delta_{F}(P^{2})=\frac{A_{d}}{2\sin d\pi}(-P^{2}-i\epsilon)^{d-2}
\label{eq:4}
\end{eqnarray}

where

\begin{equation}
A_d=\frac{16\pi^{5/2}}{{(2\pi)}^{2d}}
\frac{\Gamma(d+1/2)}{\Gamma(d-1)\Gamma(2d)}.
\label{eq:5}
\end{equation}

\begin{figure}[hbp!]
\includegraphics[height=4cm, width=7cm ]{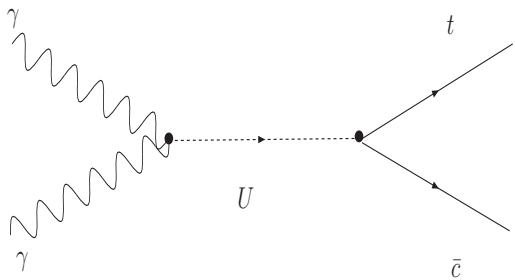}
\caption{Feynman diagram for the
$\gamma \gamma\to t \bar c$ proceses through scalar unparticle
excgange. \label{fig:0}}
\end{figure}

Since there is no tree level SM basckground effects,
the $\gamma \gamma\to t \bar c$ proceses is one of the unique processes
to probe flavor changing unparticle effects, Fig~\ref{fig:0}. The
scattering amplitude for $\gamma \gamma\to t \bar c$ process
can be written in the following

\begin{eqnarray}
\label{eq:7}
{M} = && \Big[ \epsilon^{\mu }(p_1)\big[4i
[\lambda_0(-p_1.p_2g_{\mu\nu}+p_{1\nu} p_{2\mu})
+i \lambda_0^\prime(\epsilon_{\mu\nu\alpha\beta}p_1^{\alpha}p_2^{\beta})]\big]
\epsilon^\nu(p_2) \Big]
\nonumber\\
&& \times \Big[ \bar u(p_3) [ \lambda_s-
i\gamma_5 \lambda_p ]  v(p_4)
\Big]
\big[\frac{i A_d}{2\sin{d\pi}}[-(p_1+p_2)^2]^{d-2} \big]
\big[\frac{1}{\Lambda_U^{2d-1}}\big]
\end{eqnarray}

\begin{figure}[tbp!]
\bigskip
\includegraphics{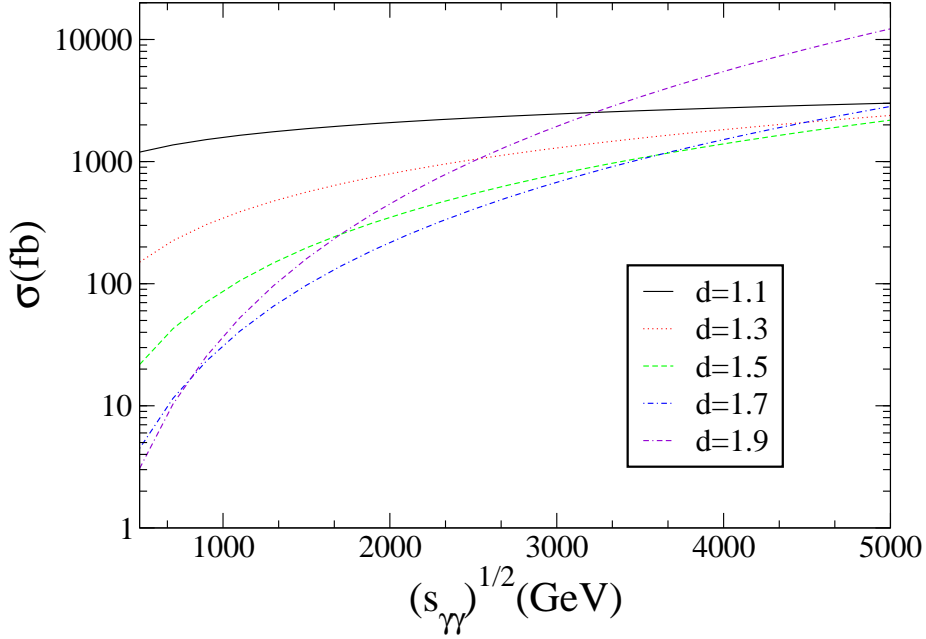}\\
\caption{The unpolarized total
cross section with  $\lambda_0/\Lambda_{\cal U}=5.10^{-1}$ TeV$^{-1}$.
\label{fig:1}}
\bigskip
\end{figure}

Using this matrix element the cross section can be calculated
straightforwardly. For the calculations of the unpolarized and polarized
cross sections, we use the expressions given in the
Appendix 1. The unpolarized total cross section
with respect to the center of mass energy of the
mono-energetic photon beams is plotted in the Figure \ref{fig:1}.
For illustration of unparticle effects, we assume $\lambda_0\equiv\lambda_s\equiv\lambda_p$
and, $\lambda_0/\Lambda_{\cal U}=0.5$ TeV$^{-1}$.

Depending on the initial electron(positron) polarization
$P_e$, and the laser beam polarization $h_l$,
the differential scattering cross section in terms of the
average helicity $h_\gamma$ can be written as

\begin{eqnarray}
\label{eq:8}
\frac{d\sigma}{d\cos\theta} = &&\frac{1}{(64\pi)}\int_{x_{1min}}^{0.83}dx_1
\int_{x_{2min}}^{0.83}dx_2
\frac{f(x_1)f(x_2)}{\hat s }
\\
&& \times \Big[ \Big ( \frac{1+ h_{\gamma}(x_1)h_{\gamma}(x_2)}{2} \Big )
\Big |M(++) \Big |^2
+\Big ( \frac{1- h_{\gamma}(x_1)h_{\gamma}(x_2)}{2} \Big )
\Big |M(+-) \Big |^2\Big]
\nonumber
\end{eqnarray}

where $f(x)=f(x,P_e,P_l)$ is the photon number density,
and $h_{\gamma}=h_{\gamma}(x,P_e,P_l)$ is the average
helicity function presented in the Appendix \ref{sec:c},
and as $\sqrt{s_{ee}}\equiv\sqrt{s}$ being the center of mass energy
of the $e^{+}e^{-}$ collider, $\sqrt{\hat s}=\sqrt{x_1x_2 s_{ee}}$
is the reduced center of mass energy of the back-scattered photon beams,
and $x=E_\gamma/E_e$ is the fraction of energy taken by
the back-scattered photon beam.

In our analysis, we follow the usual collider assumptions, and
we take $|h_l|=1$, and $|P_e|=0.9$.  Also, we use the
angular cuts $\pi/6<\cos\theta<5\pi/6$,
and $\sqrt{0.4}<x_i<x_{max}$ which have been used in the literature,
where $x_{max}$ is the maximum energy fraction of the back-scattered
photon, and its optimum value is 0.83.

\begin{figure}[tbp!]
\bigskip
\includegraphics{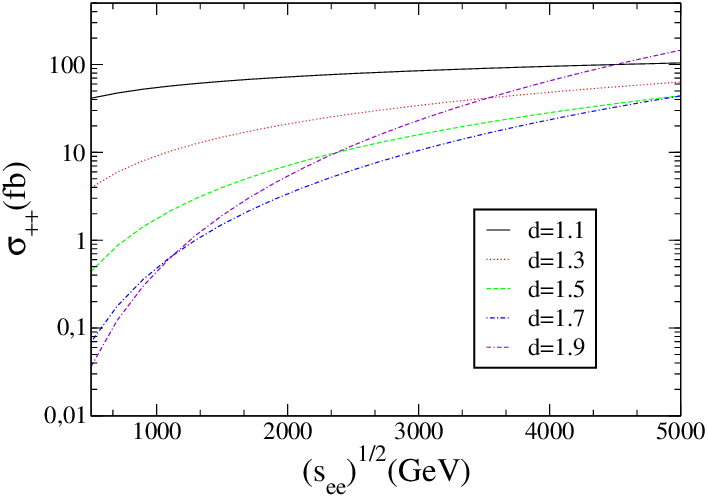}\\
\caption{The polarized cross sections with the
polarization configuration $(++)$, we assume
$\lambda_0/\Lambda_U=0.5$ TeV$^{-1}$.\bigskip }
\label{fig:2}
\bigskip
\end{figure}

\begin{figure}[tbp!]
\bigskip
\includegraphics{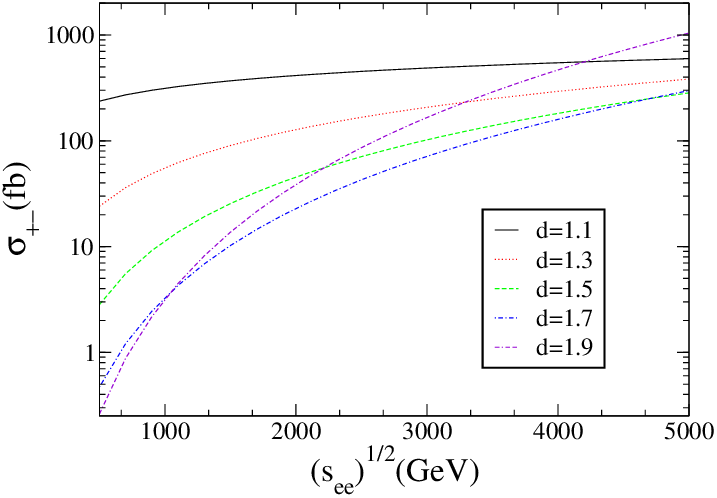}\\
\caption{The polarized cross sections with the
polarization configuration $(+-)$, we assume
$\lambda_0/\Lambda_U=0.5$ TeV$^{-1}$.\\ }
\label{fig:3}
\bigskip
\end{figure}

In Figure~\ref{fig:2}, and Figure~\ref{fig:3},
we plot the cross section for two different polarization
configurations of initial electron and laser beams
to present the behavior of
the polarized cross section with unparticle contributions
with respect to $\sqrt{s_{ee}}$.
For the figures, we use the following definitions
for the polarization configurations:
$(++)\equiv(++++)=(P_{e1}=0.9,h_{l1}=1;P_{e2}=0.9,h_{l2}=1)$,
and
$(+-)\equiv(+-+-)=(P_{e1}=0.9,h_{l1}=-1;P_{e2}=0.9,h_{l2}=-1)$, and
we take $\lambda_0/\Lambda_U=0.5$ TeV$^{-1}$.

\begin{figure}[htp!]
\bigskip
\includegraphics{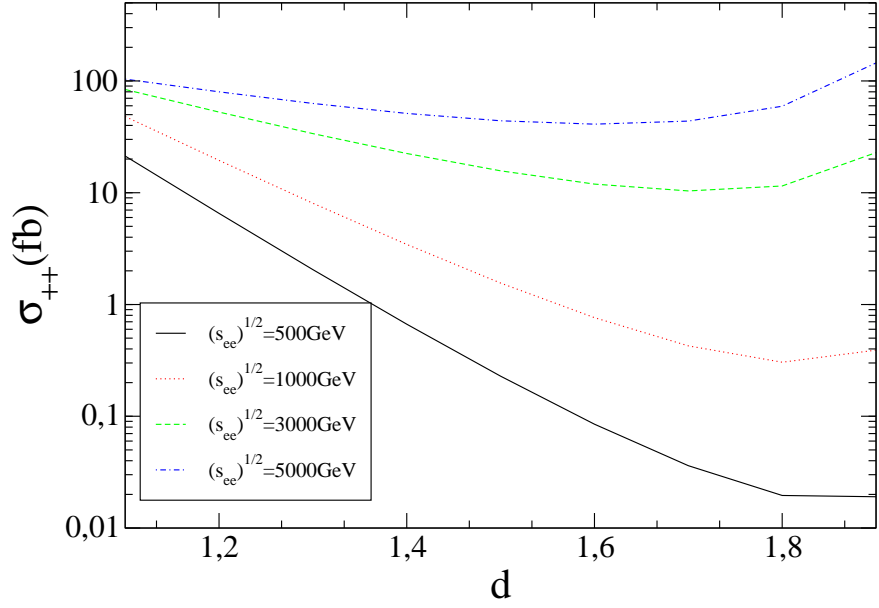}
\caption{The polarized cross sections with respect to $d$
for the polarization configuration $(++)$, we take
$\lambda_0/\Lambda_U=0.5$ TeV$^{-1}$.\\}
\bigskip
\bigskip
\label{fig:5}
\end{figure}

\begin{figure}[tbp!]
\bigskip
\includegraphics{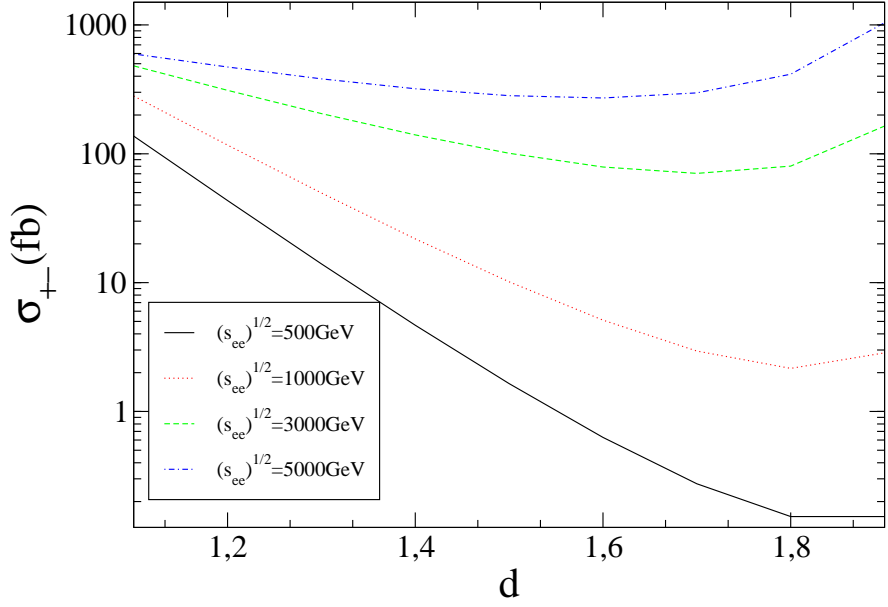}\\
\caption{The polarized cross sections with respect to $d$
for the polarization configuration $(+-)$. We assume
$\lambda_0/\Lambda_U=0.5$ TeV$^{-1}$.\\}
\label{fig:6}
\bigskip
\end{figure}

In the Figures~\ref{fig:5}, and ~\ref{fig:6}
we plot the polarized cross sections
with respect to the scaling dimension $d$
for the polarization configurations $(++)$, and $(+-)$,
respectively. From those figures one can see that for any values
of anomalous dimension $d$, the cross section $\sigma_{+-}$
is about one order larger than the $\sigma_{++}$,
i.e. to search for unparticle physics effects the
polarisation configuration $(+-)$ is more preferable
than  $(++)$.

The dependence of the cross section on
the ratio of $\sqrt{s}/\Lambda$, and $d$
can be understood in the following 
sense. From the expression of the total cross section for the
the $\gamma \gamma\to t \bar c$ process, it follows that

\begin{eqnarray}
 \sigma\sim \Lambda^{-2}(\sqrt{s}/\Lambda)^{4d-4}.
\end{eqnarray}

Therefore, since $1<d<2$ the cross section grows as $\sqrt{s}$ increases.
From Eq.~\ref{eq:7} one can easily obtain that the unitarity condition
leads to the following(in derivation this result, for simplicity,
we assume that $\lambda_s=\lambda_p$,
and $\lambda_0^\prime=0$)

\begin{eqnarray}
 \frac{\lambda_0 \lambda_s }{16\pi}\frac{A_d}{\sin (d\pi)}\frac{m_t}{\Lambda}
\Big( \frac{\sqrt{s}}{\Lambda}\Big)^{2d-2}<1.
\end{eqnarray}

Hence for the above given analysis unitarity condition is
preserved up to $\sim 100$TeV collider energies, when
$\lambda_0= \lambda_s\sim 1$ and at $\Lambda=1$TeV.

\begin{figure}[tbp!]
\bigskip
\includegraphics{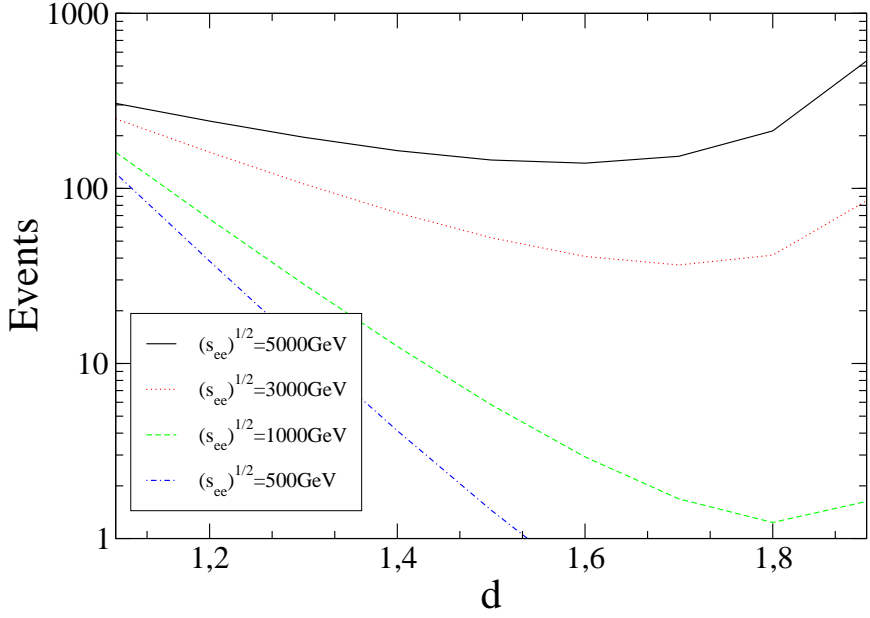}
\caption{The number of enents with respect to $d$
for the polarization configuration $(+-)$. We assume
$\lambda_0/\Lambda_U=0.1$ TeV$^{-1}$.\label{fig:7} }
\bigskip
\end{figure}

Finally, let us discuss the possility for experimental detectabilty of
the considered process. For this purpose we estimate the number of
events. As we already noted that there is no background effects for
the $\gamma \gamma\to t \bar c$ process
and we assume the number of events as the poisson variable.
In the Figure~\ref{fig:7}, we present the dependence of the predicted
number of events on the scaling dimension $d$, for the
energy options of the collider. In the analysis,
for simplicity, we take $\lambda\equiv \lambda_0=\lambda_s=\lambda_p$.
Therefore, we extract upper limits on the unparticle coupling $\lambda$
regarding $5\sigma$ analysis for the number of events
$\nu=\sigma\times {\cal L}$. For $95\%$ C.L. we take $\bar \nu \geq 9.15$.
In the Table~\ref{tab1}, and Table~\ref{tab2}
we present the limits on the unparticle coupling $\lambda$ for the photon
polarization (++), and (+-), respectively, for various $d$ values,
 and for $\Lambda=1000$GeV. As expected, from the Tables~\ref{tab1}, and ~\ref{tab2},
 one can see that with increasing $\sqrt{s}$ at given values
of $d$ the upper limits on $\lambda$ becomes more stringent.
Our limits are consistent with the limits calculated from other
low and high energy physics implications(see, for example,
\cite{unparticle, Ozansoy:2008hg}, and references there in.).

\begin{table}[tbp!]
{\caption{Upper limits on the $\lambda$ for the
polarization configuration(++++) for  ${\cal L}=1000fb^{-1}$\label{tab1}}}
\begin{ruledtabular}
\begin{tabular}{lcccccc}
$\sqrt{s}$ GeV &  d=1.1  & d=1.3 & d=1.5 & d=1.7 & d=1.9 \\
\hline
500 &$0.09 $ & $0.13$ & $0.20$ & $0.25$ & $0.30$ \\
1000 &$0.08 $ & $0.10$ & $0.15$ & $0.18$ & $0.19$ \\
3000 &$0.07 $ & $0.08$ & $0.10$ & $0.11$ & $0.10$ \\
5000 &$0.07 $ & $0.08$ & $0.08$ & $0.09$ & $0.07$ \\
\end{tabular}
\end{ruledtabular}
\end{table}

\begin{table}[tbp!]
{\caption{Upper limits on the $\lambda_0$ for the
polarization configuration(+-+-) for  ${\cal L}=1000fb^{-1}$\label{tab2}}}
\begin{ruledtabular}
\begin{tabular}{lcccccc}
$\sqrt{s}$ GeV &  d=1.1  & d=1.3 & d=1.5 & d=1.7 & d=1.9 \\
\hline
500 &$0.07 $ & $0.10$ & $0.14$  & $0.18$  & $0.20$ \\
1000 &$0.06 $ & $0.08$ & $0.10$  & $0.13$  & $0.14$ \\
3000 &$0.05 $ & $0.07$ & $0.08$ & $0.08$ & $0.07$ \\
5000 &$0.0 $ & $0.06$ & $0.06$ & $0.06$ & $0.05$ \\
\end{tabular}
\end{ruledtabular}
\bigskip
\end{table}

In conclusion, we have studied the single top
quark production in a prospected multi-TeV $\gamma\gamma$ collider
via unparticle exchange, namely the process $\gamma\gamma\to t\bar c$.
For different polarization considerations we obtain upper limits on
the flavor violating unparticle coupling constant. We show that
different polarizations give different upper limits, and as the collider
energy increases the limits gets more stringent.


\appendix

\section*{Appendix\label{sec:a}}

\subsection{}

In the calculations, we assume the following center of mass
reference frame kinematical relations

\begin{eqnarray}
p_1^{\mu}=&&E(1,0,0,1),\quad p_2^{\mu}=E(1,0,0,-1)\\
\epsilon_1^\mu=&&-\frac{1}{\sqrt{2}}(0,h_1,i,0)\quad
\epsilon_2^\mu=\frac{1}{\sqrt{2}}(0,-h_2,i,0)\\
(p_1+p_2)^2=&&s=(p_3+p_4)^2,\quad 2p_3.p_4=s-m_t^2
\end{eqnarray}

where $\epsilon_1\equiv\epsilon_1(h_1),\epsilon_2\equiv\epsilon_1(h_2)$, etc.,
$h_1,h_2=\{ +,- \}$ stand for the polarizations, $m_t$ is the mass of the
top quark.

Using these definitions we obtain

\begin{eqnarray}
|M(++)|^2=&&|M(--)|^2=8[f(d)]^2[-s]^{2d-2}
[(s-m_t^2)(\lambda_s^2+\lambda_p^2)\nonumber\\
&&-2*m_c*m_t(\lambda_s^2-\lambda_p^2)]\\
|M(+-)|^2=&&|M(-+)|^2=0
\end{eqnarray}
where

\begin{equation}
\label{a1}
f(d)=\frac{{\lambda_{0}^2} A_d}{ 2\Lambda^{2d-1}\sin (d\pi)},
\end{equation}

\subsection{\label{sec:c}}

Here, we present the definitions of the functions
appearing in the expression of the differential cross section.
Let $h_e$ and $h_l$ be the polarizations of the electron beam and
the laser photon beam, respectively. Following to
\cite{Ginzburg:1983}, 

\begin{eqnarray}
 C(x)=\frac{1}{1-x}+1-x-4r(1-r)-h_eh_lrz(2r-1)(2-x)
\end{eqnarray}

where $r=\frac{x}{z(1-x)}$. Thus, the photon number
density is given by

\begin{eqnarray}
 f(x,h_e,h_l,z)=\big(\frac{2\pi\alpha^2}{m_e^2z\sigma_c} \big)C(x)
\end{eqnarray}
where
\begin{eqnarray}
 \sigma_c=&&\big(\frac{2\pi\alpha^2}{m_e^2z\sigma_c} \big)
\big[\big(1-\frac{4}{z}-\frac{8}{z^2} \big)ln(z+1)
+\frac{1}{2}+\frac{8}{z}-+\frac{1}{2(z+1)^2} \big]\nonumber\\
&&+h_eh_l\big(\frac{2\pi\alpha^2}{m_e^2z\sigma_c} \big)
\big[\big(1+\frac{2}{z} \big)ln(z+1)-\frac{5}{2}+
\frac{1}{z+1}-\frac{1}{2(z+1)^2} \big]
\end{eqnarray}

The average helicity in terms of the function $C(x)$ can be
given by

\begin{eqnarray}
h_\gamma(x,h_e,h_l,z)=\frac{1}{C(x)}
\big\{h_e\big[\frac{x}{1-x}+x(2r-1)^2 \big]
-h_l(2r-1)\big(1-x+\frac{1}{1-x} \big) \big\}
\end{eqnarray}

\end{document}